\documentclass[twoside,twocolumn]{article}

\usepackage{blindtext} 

\usepackage[sc]{mathpazo} 
\usepackage[T1]{fontenc} 
\usepackage{microtype} 
\usepackage{amsmath,amsfonts,amssymb}
\usepackage{graphicx}
\usepackage[colorlinks=true, allcolors=blue]{hyperref}
\usepackage{siunitx}
\usepackage[english]{babel} 

\usepackage[hmarginratio=1:1,top=32mm,columnsep=20pt]{geometry} 
\usepackage[hang, small,labelfont=bf,up,textfont=it,up]{caption} 
\usepackage{booktabs} 

\usepackage{abstract} 

\usepackage{titlesec} 
\titleformat{\section}[block]{\large\scshape\centering}{\thesection.}{1em}{} 

\usepackage{fancyhdr} 
\usepackage{titling} 

\usepackage{hyperref} 
\usepackage{cite}
\newcommand\blfootnote[1]{%
  \begingroup
  \renewcommand\thefootnote{}\footnote{#1}%
  \addtocounter{footnote}{-1}%
  \endgroup
}

\newcommand{\sectionvspace}{\vspace{-3 mm}}

\pretitle{\begin{center}\Huge\bfseries} 
\posttitle{\end{center}} 
\title{\fontsize{16}{20} \selectfont BPE and computer-extracted parenchymal enhancement for breast cancer risk, response monitoring, and prognosis} 
\author{%
\textsc{Bas H.M. van der Velden, PhD}\\[1ex] 
\normalsize Image Sciences Institute, University Medical Center Utrecht, The Netherlands \\ 
\normalsize \href{mailto:bvelden2@umcutrecht.nl}{bvelden2@umcutrecht.nl} 
}
\date{} 

\begin{document}

\maketitle

\section*{Introduction}
\sectionvspace
Functional behavior of breast cancer -- representing underlying biology -- can be analyzed using MRI \cite{Esserman2006MagneticSitu,Kuhl2007CurrentApplications}. The most widely used breast MR imaging protocol is dynamic contrast-enhanced $T_1$-weighted imaging \cite{Kuhl2007The1}. The cancer enhances on dynamic contrast-enhanced MR imaging because the contrast agent leaks from the leaky vessels into the interstitial space. The contrast agent subsequently leaks back into the vascular space, creating a washout effect \cite{Hylton2006DynamicBiomarker.}.\blfootnote{This work has been distributed as part of the syllabus at the \textit{ISMRM Workshop on Breast MRI: Advancing the State of the Art}.}

The normal parenchymal tissue of the breast can also enhance after contrast injection. This enhancement generally increases over time \cite{Giess2014BackgroundInterpretation.}. Typically, a radiologist assesses this background parenchymal enhancement (BPE) using the Breast Imaging Reporting and Data System (BI-RADS) \cite{Morris2013ACRImaging}. According to the BI-RADS, BPE refers to the volume of enhancement and the intensity of enhancement and is divided in four incremental categories: minimal, mild, moderate, and marked.

Researchers have developed semi-automatic and automatic methods to extract properties of BPE from MR images. For clarity, in this syllabus the BI-RADS definition will be referred to as BPE, whereas the computer-extracted properties will not.

Both BPE and computer-extracted parenchymal enhancement properties have been linked to screening and diagnosis, hormone status and age, risk of development of breast cancer, response monitoring, and prognosis. 


\section*{Background parenchymal Enhancement}
\sectionvspace
Giess et al. gave an exhaustive overview of the appearance of BPE \cite{Giess2014BackgroundInterpretation.}. BPE has been shown to be lower in older and in postmenopausal women and patients \cite{Chen2013BackgroundDCE-MRI,Bennani-Baiti2016MRICancer,Sogani2017ComparisonImaging,You2018DecreasedCancer}.

In screening, MR images with high levels of BPE were more likely to receive an abnormal interpretation, which might lead to recommendations for further testing \cite{Hambly2011BackgroundFollow-up,DeMartini2012BackgroundPerformance,Ray2018EffectPractices}. Increased BPE has also been associated with inaccurate tumor size estimation and inaccurate tumor staging \cite{Uematsu2011DoesCancer,Baek2014BackgroundEstimation.}.

High BPE shows potential as a biomarker for risk assessment of developing breast cancer \cite{King2011BackgroundRisk.,DeLeo2015BreastSalpingooophorectomy.,Dontchos2015AreRisk,Telegrafo2016BreastCancer,Melsaether2017BackgroundCancer}. However, in women without a BRCA mutation, this appeared to be confounded by age \cite{Bennani-Baiti2016MRICancer}. In patients in whom a cancer is detected, those with low BPE had a higher grade tumor which is more likely to be progesterone negative \cite{Vreemann2018TheCancers.}.

BPE generally decreases after systemic treatment of breast cancer. This decrease in BPE was found for anti-hormonal therapy \cite{Sogani2017ComparisonImaging,King2012EffectImaging,King2012ImpactImaging,Mousa2012TheTrial.,Kim2017DiagnosticTherapy}, chemotherapy \cite{Kim2017DiagnosticTherapy,Chen2013BackgroundDCE-MRI}, and radiation therapy \cite{Sogani2017ComparisonImaging}.

A substantial agreement was found between BPE assessment on dynamic contrast-enhanced MRI and its counterpart on contrast-enhanced mammography \cite{Sogani2017ComparisonImaging}. BPE also correlated with uptake of $^{\text{99m}}$Tc-methoxy isobutyl isonitrile in the breast \cite{Yoon2015BackgroundImaging.}. BPE did, however, not appear to be related to $^{\text{18}}$F-fludeoxyglucose uptake on positron emission tomography \cite{Koo2013BackgroundMRI}.


\section*{Computer-extracted parenchymal enhancement}
\sectionvspace
Computer-extracted features of parenchymal enhancement are typically generated using the following workflow: The parenchymal tissue is segmented on the MR images after which features are calculated using these segmented voxels.

In semi-automatic methods, several approaches exist. Examples include manual delineation of parenchymal tissue on the MR images by an expert and growing of parenchymal tissue segmentations from user-selected seed points. In automatic methods, computer algorithms perform all the steps necessary to get to a parenchymal tissue segmentation (Figure 1).

\begin{figure*}
  \includegraphics[width=\textwidth]{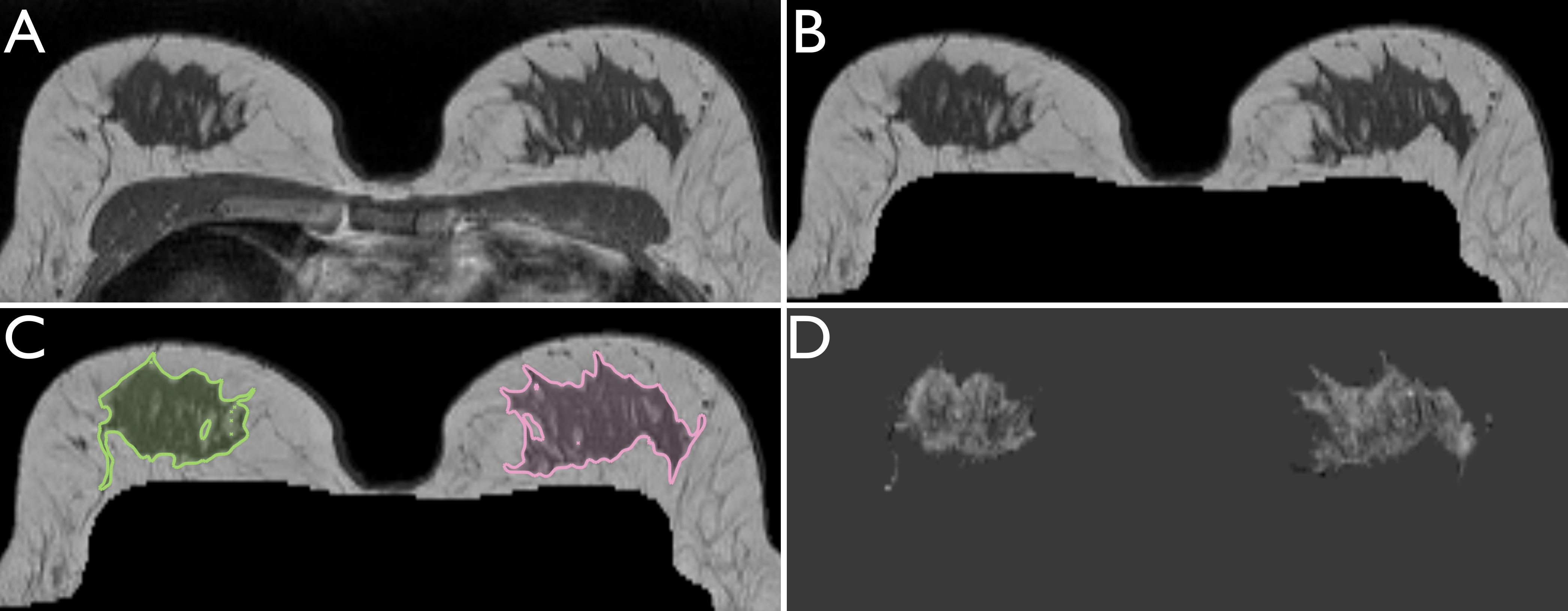}
  \caption{Example of an automatic image analysis pipeline for calculation of parenchymal enhancement features. A: Transversal pre-contrast breast MRI scan. B: The breast area is automatically segmented. C: The parenchymal tissue in each breast is segmented. D: The enhancement of the parenchymal tissue between the pre-contrast scan and the first post-contrast scan. On these voxels, features are typically calculated.}
\end{figure*}

Once the parenchymal tissue is segmented, numerous features can be calculated. Some common features include the percent enhancement, the signal enhancement ratio, or texture features.

Analogous to BPE, these computer-extracted features can serve as biomarker for the risk of developing breast cancer. They often are often used, however, as biomarkers for different tasks such as tumor recurrence, response to treatment, or patient survival. Examples of these biomarkers and their applications can be found in Table 1.

\section*{Example: CPE as biomarker of survival}
\sectionvspace
An example of such a biomarker from our group is contralateral parenchymal enhancement (CPE) \cite{vanderVelden2015AssociationCancer}. In a cohort of 398 patients with estrogen receptor (ER)-positive/HER2-negative breast cancer from the Netherlands Cancer Institute in Amsterdam, we found that patients who have high enhancement in the parenchyma of the contralateral breast (i.e. high CPE) have a significantly better survival than patients with low CPE. This was independent of potential confounding variables such as patient age and tumor pathology markers \cite{vanderVelden2015AssociationCancer}. 

The CPE biomarker was validated in an independent cohort of 302 patients with ER-positive/HER2-negative breast cancer from the Memorial Sloan Kettering Cancer Center in New York \cite{vanderVelden2018ContralateralPatients}. 

In patients considered to be at high-risk by molecular assays such as the 70-gene signature and the 21-gene recurrence score, CPE could identify a group of patients at relatively low risk \cite{vanderVelden2017ComplementaryCancer}.

\section*{Outlook}
\sectionvspace
BPE and computer extracted features of parenchymal enhancement show substantial potential as biomarkers for breast cancer of diagnosis, response monitoring, prognosis, and many other clinically relevant tasks.  There are, however, still several opportunities to further progress the field, especially for the computer-extracted biomarkers. Before these can be used at large scale, some issues should be addressed.

The computer-extracted biomarkers should be made agnostic to deviations to MR imaging vendors and protocols, since imaging parameters could influence the biomarkers. Image analysis methods must produce similar results across vendors and protocols. Recent advances in machine learning and most notably deep learning show potential for this. For example, in a recent MICCAI challenge on brain white matter lesion segmentation, deep learning algorithms showed excellent performance on images from scanners that the algorithms were not trained on (\url{http://wmh.isi.uu.nl}). 

The biology of these computer-extracted biomarkers should be better understood, e.g. by analyzing radiogenomic or radioproteomic datasets. These 'multi-omic' datasets can even be used for better prediction of disease outcome. An example of this approach combined MR imaging features of tumor and parenchyma with genomic data to find multidimensional clusters related to patient survival \cite{Wu2017UnsupervisedPathways.}.

\section*{Conclusion}
\sectionvspace
BPE and computer extracted features of parenchymal enhancement show potential as biomarkers for clinically relevant tasks.

\begin{table*}[]
\caption{Examples of computer-extracted parenchymal enhancement biomarkers. NB This list is not meant to be exhaustive.}
\footnotesize
\begin{center}
\begin{tabular}{llll}
\hline
Authors                        & Year & Parenchyma features              & Clinical endpoint                                              \\
\hline
Aghaei et al. \cite{Aghaei2015Computer-aidedChemotherapy}         & 2015 & Bilateral PE                     & Complete reponse scored on RECIST                              \\
Fan et al. \cite{Fan2017RadiomicPatients}            & 2017 & Bilateral PE                     & Complete reponse scored on RECIST                              \\
Hattangadi et al. \cite{Hattangadi2008BreastChemotherapy}     & 2008 & SER surrounding the tumor        & Disease-free survival after NAC                            \\
Jones et al.  \cite{Jones2013MRIChemotherapy}          & 2013 & PE surrounding the tumor         & Disease-free survival after NAC                            \\
Kim et al. \cite{Kim2013PredictingMRI}           & 2013 & SER surrounding the tumor        & Ipsilateral breast tumor recurrence                            \\
Knuttel et al. \cite{Knuttel2016PredictionCancer}        & 2016 & SER surrounding the tumor        & Extensive ductal carcinoma in situ                             \\
Luo et al. \cite{Luo2017DuctalTreatment}           & 2017 & Functional Tumor Volume, SER, PE & DCIS recurrence after treatment            \\
Nabavizadeh et al. \cite{Nabavizadeh2011TopographicMRI.}    & 2011 & PE surrounding the tumor         & Microvessel density, genomic changes                           \\
van der Velden et al. \cite{vanderVelden2015AssociationCancer} & 2015 & Contralateral PE                 & Overall and disease-free survival                              \\
van der Velden et al. \cite{vanderVelden2018ContralateralPatients} & 2018 & Contralateral PE                 & Overall and disease-free survival validation                   \\
van der Velden et al. \cite{vanderVelden2017ComplementaryCancer} & 2017 & Contralateral PE                 & Complementary value to 70-GS and 21-GRS                          \\
Wang et al. \cite{Wang2015IdentifyingStudy.}           & 2015 & Density, enhancement, texture    & Triple-negative status tumor                                   \\
Wu et al. \cite{Wu2015QuantitativeCarriers}    & 2015 & FGT volume and PE                & Breast cancer risk after RRSO   \\
Wu et al. \cite{Wu2016BreastCancer.}             & 2016 & Wash-in slope variance and SER   & Breast cancer risk     \\
You et al. \cite{You2017AssociationPatients}            & 2017 & Contralateral PE change          & Pathological complete remission after NAC\\
\hline

\end{tabular}
\end{center}
70-GS = 70-gene signature, 21-GRS = 21-gene recurrence score, DCIS = ductal carcinoma in situ, FGT = fibroglandular tissue, NAC = neoadjuvant chemotherapy, PE = percent enhancement, RECIST = response evaluation criteria in solid tumors, RRSO = risk-reducing salpingo-oophorectomy, SER = signal enhancement ratio
\end{table*}


\footnotesize
\bibliographystyle{ieeetr}
\bibliography{}

\end{document}